\documentclass[conference]{IEEEtran}
\IEEEoverridecommandlockouts

\usepackage{amsmath,amssymb,amsfonts}
\usepackage{algorithmic}
\usepackage{graphicx}
\usepackage{textcomp}
\usepackage{threeparttable}
\usepackage{optidef}
\usepackage{xcolor}
\usepackage[utf8]{inputenc}
\DeclareUnicodeCharacter{2082}{\textsubscript{2}}

\def\BibTeX{{\rm B\kern-.05em{\sc i\kern-.125em b}\kern-.08em
    T\kern-.1667em\lower.7ex\hbox{E}\kern-.125emX}}
    
\usepackage[
backend=biber,
style=ieee,
sorting=none
]{biblatex}

\begin{document}

\title{Optimal Constant Climb Airspeed with Variable Cost Index for All-electric Aircraft\\
}

\author{\IEEEauthorblockN{Lucas Souza e Silva and Luis Rodrigues}
\IEEEauthorblockA{\textit{Department of Electrical and Computer Engineering} \\
\textit{Concordia University}\\
Montreal, Canada 
}

}
\maketitle

\begin{abstract}
This paper presents for the first time an approach to minimize direct operational costs (DOC) for all-electric aircraft during the climb phase, introducing a time-varying cost index (CI). The CI is modeled as a dynamic parameter commanded by Air Traffic Control (ATC), allowing the aircraft to maintain a constant airspeed throughout the climb, while respecting the air traffic regulations. This paper also explores the implications of a time-varying CI on the determination of optimal airspeed and climbing time for all-electric aircraft. Additionally, it provides the necessary equations to calculate both the optimal climb airspeed and climb duration. The proposed methodology has been validated through a simulated scenario that reflects actual operational procedures. As a result, optimal values for climb airspeed, climbing time, and energy consumption have been established, paving the way for future applications of this methodology to advanced air mobility all-electric vehicles.
\end{abstract}

\section{Introduction}\label{sec1}
The interest for fully electric solutions is continuously increasing in the aviation community, especially in the past few years, due to their potential profitability and capability to emerge as forefront responses to greenhouse gas (GHG) emissions. The global electric aircraft market was valued at USD 7.91 billion in 2022 and it is projected to reach USD 50.86 billion by 2032 \cite{PrecResearch2023}. From an environmental perspective, the aviation sector accounts for approximately 2.5\% of the world{'}s current $CO_{2}$ emissions \cite{Ritchie2024}. At a first glance, it does not seem alarming, but only a small portion of the global population has access to aviation services. This indicates the potential impact in carbon dioxide emissions caused by aviation worldwide in a scenario of steadly increasing demand for domestic and international travel. The rapid urbanization of populated areas, the expanding awareness with new environmental concerns and the fast-paced rise in air travel demand are propelling initiatives that promote technological advance in transportation. Examples of such developments include autonomous systems and reliable alternatives to fossil-based fuels. In particular, the Advanced Air Mobility (AAM) concept intends to provide transportation services for people and cargo in an automated and cooperative fashion \cite{FAA_Conops}.

One of the main technical challenges faced by all-electric aviation is the limited energy density of the electrical batteries \cite{Barzkar_Ghassemi2022}, which restricts their operation in long-haul flights. One alternative for improving energy efficiency in aviation is to minimize overall direct operational costs (DOC) by flying the aircraft in the optimal airspeed that corresponds to the economy (ECON) mode as part of the flight plan in modern aircraft Flight Management Systems (FMS). The ECON speed is defined for any flight phase based on the cost index ($CI$), which is a trade-off parameter used in aviation characterized by the ratio of the costs associated to time of operation (crew salaries, maintenance procedures, costs incurred by prolonged delays or leasing of equipment) and the cost due to energy consumption (the cost to charge electrical batteries for all-electric aircraft). The minimization of DOC has been an active research topic since the introduction of FMS in the early 1980s. Initial research on the determination of optimal airspeed can be found in \cite{Erzberger_Lee1980, Liden1985}. More recent work that deals with minimizing DOC typically uses optimal control theoretical results, such as the Hamilton-Jacobi-Bellman (HJB) equation or the Pontryagin{'}s Minimum Principle (PMP) to find optimal and suboptimal airspeed for fuel-powered aircraft \cite{Villarroel_Rodrigues2016a, Fan_et_al2020, Villarroel_Rodrigues2016b} and for all-electric aircraft \cite{Kaptsov_Rodrigues2017, Wang2020}. However, these references assume a constant $CI$ throughout the flight phase for which the optimal airspeed has been computed. Some research considers $CI$ as a parameter that might change during flight \cite{Cook_et_al2009, Camacho_et_al2015, Prats_Torre_Delgado2022, Oliveira_et_al2023, Mori2022}. In \cite{Silva_Akgunduz_Rodrigues2024}, the authors introduce a time-varying $CI$ that balances the strategic objectives of airlines with operational restrictions imposed by Air Traffic Control (ATC) as part of the cruise ECON speed computation. Nevertheless,  to the best of the authors{'} knowledge, none of the previous work in the open literature shows how ATC-driven changes in $CI$ affect the computation of the optimal climb airspeed and climbing time for all-electric aircraft. This paper contrasts with the existing contributions as it demonstrates how a time-varying $CI$ impacts the determination of the optimal climb airspeed and climbing time of all-electric aircraft for operations that require climbing in constant airspeed.
The main contributions of this paper are:
\begin{enumerate}
\item The introduction of $CI$ as a dynamic parameter in the formulation of the optimization problem to minimize DOC for all-electric aircraft in climb. The changes in $CI$ are commanded by ATC to impose operational restrictions to the all-electric aircraft operation.
\item The equations that compute the optimal climb airspeed and time are provided for all-electric aircraft. The aircraft energy consumption is also determined in this paper.
\item The validation of the proposed methodology using a climb profile inspired by an actual operational procedure based on regulatory standards.
\end{enumerate}

Several operational procedures require constant airspeed climb, such as Noise Abatement Departure Procedures (NADP) and specific Climb Via Clearances. NADPs were created to alleviate the effects of noise caused by aircraft operation in terminal areas to communities that live nearby \cite{FAA_Noise_Ab1993}, by creating flight profiles that reduce exposure of individuals on the ground to the noise primarily caused by the aircraft{'}s engines. According to \cite{ICAO_Noise_Ab2007}, most of NADPs require constant airspeed during the majority of the aircraft climb phase. Climb Via Clearance procedures implement ATC requirements to make the aircraft climb using specific paths or waypoints with restrictions on altitude and speed. An example of a requirement of compliance with ATC clearance is the FAR 14 CFR 91.123 in the United States \cite{14CFR_ATC_Clear}. In accordance with procedures that require constant airspeed in climb, \cite{FAA_IFH2012} provides pilots with instructions on how to enter and maintain constant speed climb maneuvers. Although these procedures were initially proposed for fuel-powered aircraft, they could be adapted to all-electric aircraft, particularly within the controlled airspace, once these vehicles are fully integrated to the airspace system. 

This paper is organized as follows: Section II presents the methodology to perform the FMS initialization and the calculation of optimal climb airspeed and climbing time with variable $CI$ for all-electric aircraft. Section III presents a simulated scenario and discussions about the observed results. Section IV concludes the paper.

\section{Problem Statement and Solution}\label{sec3}
\subsection{Aircraft Dynamic Model and Assumptions}\label{subsec3.A}

Let us consider that $x$ describes the horizontal position of the aircraft, $h$ is the aircraft vertical position (altitude), $v$ is its airspeed, $D$ is the magnitude of the drag force, $L$ is the magnitude of the lift force, $T$ is the magnitude of the aircraft’s thrust force, $W$ is the aircraft’s weight, $\gamma$ is the flight path angle and $CI$ is the aircraft cost index. The following assumptions are made:

\begin{enumerate}
\item The flight Mach number is assumed to be below the drag divergence Mach number, with the aircraft operating within its flight envelope.
\item The aircraft is assumed to be a fixed-wing aircraft, so the wing surface area $S$ is constant.
\item The flight path angle $\gamma$ of the aircraft trajectory is positive $(\gamma > 0)$ for take-off/climb. Also, an approximation for small angles will be considered for $\gamma$. Thus, $cos\gamma \approx 1$ and $sin\gamma \approx \gamma$.
\item The battery system that provides energy to the all-electric aircraft is considered ideal, with neglectable internal resistance and it operates in cruise with constant voltage $U$. As a result, $\frac{d}{dt}(QU)=\dot Q U$.
\item The aircraft flight is steady, with no winds. As a consequence, the aircraft acceleration is neglected. Additionally, inspired by operational procedures that require constant airspeed climb as seen in section \ref{sec1}, $v$ is assumed to be constant during climb and it is aligned with the thrust force $T$.
\item The aircraft’s climb rate $\dot{h}$ is constant and equal to the average climb rate $\bar{\dot{h}}$ known from historical data or operational procedures, such as the Standard Instrument Departure (SID) procedures. An example of SID for Pierre Elliot Trudeau International Airport in Montreal, Canada (IATA: YUL, ICAO: CYUL) can be found in \cite{Jeppesen_YUL2013}. 
\end{enumerate}

In \cite{Silva_Akgunduz_Rodrigues2024}, a time-varying $CI$ was introduced as the output of a first-order filter defined by

\begin{equation}
    \tau\dot{CI}=-CI+CI_{in}\label{eq_filt}
\end{equation}
where $CI_{in}$ is the forcing term that represents an input received by ATC and the initial condition $CI(0) = CI_{0}$ is the initial cost index defined by the airline or operator of the aircraft based on their strategy. In general, $CI$ is given in units of energy per units of time and in this paper, $CI$ is determined in $kJ.s^{-1}$.

All-electric aircraft energy is sourced by its batteries and its total electric charge is denoted as $Q$. From the definition of efficiency in the conversion of the electrical power to mechanical power, the electrical current $i$ supplied by the aircraft battery can be expressed as
\begin{equation}
-i=\dot{Q}=-\frac{Tv}{\eta U}\label{eq_i}
\end{equation}
where $\eta$ is the electrical system efficiency coefficient. The electrical energy for a battery system of constant voltage $U$, as per assumption 4) is given by
\begin{equation}
    E = QU \label{eqElec_energy}
\end{equation}

From Newton{'}s second law, the sum of the components of the forces acting on the aircraft while climbing with constant airspeed along the flight path can be expressed as

\begin{equation}
T - D - Wsin\gamma = 0 \label{eq_Newton}
\end{equation}

Using the approximation for small angles as per assumption 3), (\ref{eq_Newton}) is equivalent to 

\begin{equation}
T - D - W\gamma = 0 \label{eq_Newton_approx}
\end{equation}

The aircraft airspeed $v$ and climb rate $\dot{h}$ can be respectively written as 
\begin{equation}
\dot{x} = vcos\gamma \approx v \label{eq_vec_x}
\end{equation}
\begin{equation}
\dot{h} = vsin\gamma \approx v\gamma\label{eq_vec_h}
\end{equation}

Using the result from (\ref{eq_vec_h}) in (\ref{eq_Newton_approx}) yields

\begin{equation}
    \dot{h} = \frac{(T-D)v}{W}\label{eq_h_dot}
\end{equation}

As per assumption 1), the aircraft operates below the drag divergence Mach number, and assuming that it follows a drag polar curve, the magnitude of the drag force is

\begin{equation}
D=\frac{1}{2}\rho S C_{D,0}v^{2} + \frac{2C_{D,2}W^{2}}{\rho S v^{2}}\label{eq_D}
\end{equation}
where $C_{D,0}$ is the parasitic drag coefficient at zero-lift, $C_{D,2}$ is the drag coefficient induced due to lift and $\rho$ is the air density.
\\

\textit{Earth{'}s atmosphere model:}
The standard atmosphere model developed by the NASA Glenn Research Center \cite{NASA_Atm} will be used in this paper to describe the variation of the air density $\rho$ as a function of the aircraft altitude $h$ for the Troposphere atmospheric layer as

\begin{equation}
    \rho = 4.1748e^{-11}(288.14 - 0.00649h)^{4.256} \label{rho_troposphere}
\end{equation}

\subsection{Problem Formulation}\label{subsec3.B}
The direct operating cost (DOC) of an aircraft in climb is
\begin{equation}
DOC=\int_{0}^{t_{c}} (C_{t}-C_{e}\dot{Q}) \,dt\label{eq_DOC}
\end{equation}
where $t_{c}$ is the total climbing time from the waypoint usually considered at the takeoff position to the waypoint where the cruise starts. This notion can be expanded for a climb composed of several segments, where $t_{c}$ indicates the total climbing time in each of these segments. Assuming that the cost of energy $C_{e}$ is positive, one can divide (\ref{eq_DOC}) by $C_{e}$, resulting in the the cost function $J$ as
\begin{equation}
J = \frac{DOC}{C_{e}} = \int_{0}^{t_{c}} (CI -\dot{Q}) \,dt\label{eq_J}
\end{equation}
where $\frac{C_t}{C_e} = CI$ is the cost index. The minimization of DOC for an all-electric aircraft climbing with constant airspeed and variable cost index can be formulated as an optimal control problem as
 
\begin{equation}
\begin{aligned}
J^{*} = \min_{v,t_{c}} \quad & \int_{0}^{t_{c}} (CI -\dot{Q}) \,dt \label{eq_opt_prob}\\
\textrm{s.t.} \quad & \dot{x}=v\\
  & \dot{h} = \frac{(T-D)v}{W} \\
  & \dot{Q}=-\frac{Tv}{\eta U}\\
  &\tau\dot{CI}=-CI+CI_{in}\\
  &D=\frac{1}{2}\rho S C_{D,0}v^{2} + \frac{2C_{D,2}W^{2}}{\rho S v^{2}}\\
  &CI(0)=CI_{0}, Q(0)=Q_{0}\\
  &x(0)=x_{0}, x(t_{c})=x_{c}\\
  &h(0)=h_{0}, h(t_{c})=h_{c}\\
  &v>0
\end{aligned}
\end{equation}
where $J^{*}$ is the minimum DOC of an all-electric aircraft in climb achieved for the minimizers of (\ref{eq_opt_prob}), which are the optimal climb airspeed $v^{*}$ and the optimal climbing time $t_{c}^{*}$. The initial condition of the cost index $CI_{0}$ is selected by the airline for the FMS initialization when preparing the aircraft flight plan. ATC determines the flight level $h_{c}$ for the aircraft cruise. Changes in the aircraft{'}s $CI$ are expected throughout the climb from ATC to adjust the air traffic flow in a certain airspace or to conform with operational procedures, such as the NADPs. The magnitude of the step change in $CI$, noted as $CI_{in}$ is a result of multiple factors that depend on environmental, situational, or operational conditions. Therefore, we assume that $CI_{in}$ is also provided by ATC to the pilots along with the cruise flight level $h_{c}$.

\textit{Remark:} As the climb phase is typically shorter than cruise, we assume that the aircraft scheduling, including the management of delays or anticipations will be performed during cruise \cite{Silva_Akgunduz_Rodrigues2024}. 

\subsection{Problem Solution}\label{subsec3.C}
Based on problem assumption 5), the aircraft climbs at constant airspeed. For any $v\neq0$, the total climbing time $t_{c}$ can be expressed as
\begin{equation}
t_{c} = \frac{\sqrt{(x_{c}-x_{0})^2 + (h_{c}-h_{0})^2}}{v} = \frac{d}{v}\label{eq_tc}
\end{equation}

The solution of (\ref{eq_filt}), for the initial condition $CI(0)=CI_{0}$ and input $CI_{in}$, is given by
\begin{equation}
CI(t)=e^{-\frac{t}{\tau}}(CI_{0}-CI_{in})+CI_{in} \label{eq_CI}
\end{equation}
where $\tau$ is the time constant of the first-order filter and indicates the convergence rate of the $CI$ to reach the commanded value $CI_{in}$. For a time-varying $CI$ as per (\ref{eq_CI}), one can rewrite the total cost function $J$ from (\ref{eq_opt_prob}) using (\ref{eq_tc}) and (\ref{eq_CI}) as

\begin{equation}
J=\tau(CI_{0}-CI_{in})(1-e^{-\frac{d}{\tau v}})+CI_{in} \frac{d}{v}+Q_{0} -Q_{f}\label{eq_J_ext}
\end{equation}
where $Q(t_{c})=Q_{f}$ is the final battery charge when the aircraft finalizes the climb phase. Applying the necessary condition for optimality yields

\begin{equation}
\frac{\partial J}{\partial v} = -\frac{(CI_{0}-CI_{in})d}{v^{2}}e^{-\frac{d}{\tau v}} -CI_{in}\frac{d}{v^{2}}-\frac{\partial Q_{f}}{\partial v}=0\label{eq_dJ}
\end{equation}

The optimal airspeed $v^{*}$ for a variable $CI$ is the solution of (\ref{eq_dJ}), for any finite $\tau>0$ and $d>0$. 

From (\ref{eq_h_dot}), the magnitude of the thrust force $T$ can be expressed as
\begin{equation}
    T = \frac{W\dot{h}}{v} + D \label{eq_thrus}
\end{equation}

Replacing (\ref{eq_D}) in (\ref{eq_thrus}) and then (\ref{eq_thrus}) in (\ref{eq_i}) yields

\begin{equation}
\dot{Q} = \frac{-1}{\eta U} \Biggl(W\dot{h} + \frac{1}{2}\rho SC_{D,0}v^3 + \frac{2C_{D,2}W^2}{\rho Sv}\Biggl) \label{eq_Q_dot_exp}
\end{equation}

The solution of (\ref{eq_Q_dot_exp}), which is a separable differential equation, leads to

\begin{equation}
\begin{aligned}
    \int_{Q{0}}^{Q_{f}}\,dQ = \frac{-1}{\eta U} \Biggl(W  \int_{0}^{t_{c}}\,\dot{h} dt + \frac{SC_{D,0}v^3}{2}  \int_{0}^{t_{c}}\,\rho dt + \\ \frac{2C_{D,2}W^2}{Sv} \int_{0}^{t_{c}}\,\delta_{\rho} dt  \Biggl) \label{eqQ_dot_int}
    \end{aligned}
\end{equation}
where $\delta_{\rho} = \rho^{-1}$. As seen in (\ref{rho_troposphere}), $\rho$ is a function of the altitude that the aircraft is flying at and the aircraft{'}s altitude is a function of time. Then, to solve (\ref{eqQ_dot_int}) we will use the Mean Value Theorem. Let $f$ be a continuous and bounded function on a closed interval. The Mean Value Theorem states that 

\begin{equation}
    \int_{t_{a}}^{t_{b}}\,f(\lambda) d \lambda = (t_{b} - t_{a}) \bar{f} \label {eqmean_value_theorem}
\end{equation}
where $\bar{f}$ is the average value of the function $f$. Applying (\ref{eqmean_value_theorem}) in (\ref{eqQ_dot_int}) yields

\begin{equation}
    Q_{f} = Q_{0} - \frac{d}{\eta U} \Biggl(\frac{W \bar{\dot{h}}}{v} + \frac{\bar{\rho}SC_{D,0}v^2}{2} + \frac{2C_{D,2}W^2\bar{\delta_{\rho}}}{Sv^2} \Biggl) \label{eqQ_f}
\end{equation}
where $\bar{\dot{h}}$ is the mean value of the climb rate $\dot{h}$, that can be extracted from historical data or defined by operational requirements and $\bar{\rho}$ and $\delta_{\rho}$ are the mean value of $\rho$ and $\rho^{-1}$, respectively, that can be computed using (\ref{rho_troposphere}), knowing $h_c$. With the result (\ref{eqQ_f}), one can obtain the total cost function $J$ for the FMS operating in climb with an ATC input as per (\ref{eq_J_ext}). Also, from (\ref{eqQ_f}), we can now compute $\frac{\partial Q_{f}}{\partial v}$ as

\begin{equation}
\frac{\partial Q_{f}}{\partial v} = - \frac{d}{\eta U} \Biggl(-\frac{W \bar{\dot{h}}}{ v^2} + \bar{\rho}SC_{D,0}v - \frac{4C_{D,2}W^2\bar{\delta_{\rho}}}{Sv^3}\Biggl)\label{eqdE_dv_elec}
\end{equation}

Replacing (\ref{eqdE_dv_elec}) in (\ref{eq_dJ}), one can compute the optimal airspeed for the aircraft climb operation under ATC input for an all-electric aircraft. It is also noteworthy that for a constant altitude operation, which is an assumption commonly made for cruise, the air density is also constant and $\dot{h} = 0$. In this case, (\ref{eqdE_dv_elec}) becomes the same equation as presented in \cite{Silva_Akgunduz_Rodrigues2024}.

\subsection{Sufficient Condition for optimality}\label{subsecE}

To confirm that the optimal airspeed $v^*$ is a minimizer of the total cost function $J$ in (\ref{eq_J_ext}), the sufficient condition for optimality (\ref{eq_suf_cond}) shall be satisfied.

\begin{equation}
\begin {aligned}
\frac{\partial^{2} J}{\partial v^{2}} = \frac{(CI_{0}-CI_{in}) d e^{-\frac{d}{\tau v}}}{v^{4}} \Biggl(2v-\frac{d}{\tau}\Biggl) + \frac{2CI_{in}d}{v^{3}} + \\ \frac{d}{\eta U} \Biggl(\frac{W \bar{\dot{h}}}{v^3} + \bar{\rho}SC_{D,0} + \frac{12C_{D,2}W^2\bar{\delta_{\rho}}}{Sv^4}\Biggl) > 0 \label{eq_suf_cond}
\end {aligned}
\end{equation}

\subsubsection{FMS Initialization}\label{subsubsec1} 
If no ATC input is received throughout the climb, the aircraft should operate with the fixed value of $CI=CI_{0}$, which is the $CI$ value defined by the airline based on its strategy. In this case, a particular solution can be derived for the FMS initialization by making $\tau=\infty$ in (\ref{eq_dJ}), which results in
\begin{equation}
-CI_{0}\frac{d}{v_{0}^{2}}-\frac{\partial Q_{f}}{\partial v_{0}}=0\label{eq_v0}
\end{equation}
where $v^{*}_{0}$ is the optimal climb airspeed computed for the FMS initialization and $\frac{\partial Q_{f}}{\partial v_{0}}$ is given by (\ref{eqdE_dv_elec}) with $v=v_{0}$. The optimal climbing time $t^{*}_{c_{0}}$ can be computed using (\ref{eq_tc}) with $v=v^{*}_{0}$.

\section{Results and Discussions}\label{sec4}
\subsection{Simulation Parameters} \label{subsec4a}
The simulations presented herein were performed in MATLAB installed on a laptop equipped with 16 GB of RAM and an $11^{th}$ Gen Intel(R) Core(TM) i5-1135G7 2.40GHz CPU. To simulate an all-electric aircraft, we use data from a Yuneec International E430 two-seater aircraft model \cite{Li_Rodrigues2023}, \cite{E430}, as per Table \ref{table_par}.

\begin{table} [hbt]
\centering
  \begin{threeparttable}
   \caption{Simulation Parameters}\label{table_par}
   \centering
   \begin{tabular}{|c c |}
   \hline
     \textbf{Parameter} & \textbf{Value}\\
     \hline
         Wing surface area $S$ $(m^{2})$ & 11.37 \\ 
         Aircraft Mass $(kg)$ & 472\tnote{1} \\
         Zero-lift drag coefficient $C_{D,0}$ & 0.035 \\
         Induced drag coefficient $C_{D,2}$ & 0.009\\
         Maximum airspeed $v_{max}$ $(km/h)$ & 161\\
         Battery Output Voltage $U$ $(V)$ & 133.2\\
         Electrical system efficiency $\eta$ & 0.7\\ [1ex] 
     \hline
     \end{tabular}
     \begin{tablenotes}
       \item [1] Aircraft maximum take-off mass, which is constant throughout the flight.
     \end{tablenotes}
  \end{threeparttable}
\end{table}

\subsection{Simulated flight scenario}\label{subsec4b}
\begin{figure}[hbt!]
\centerline{\includegraphics[scale=0.24]{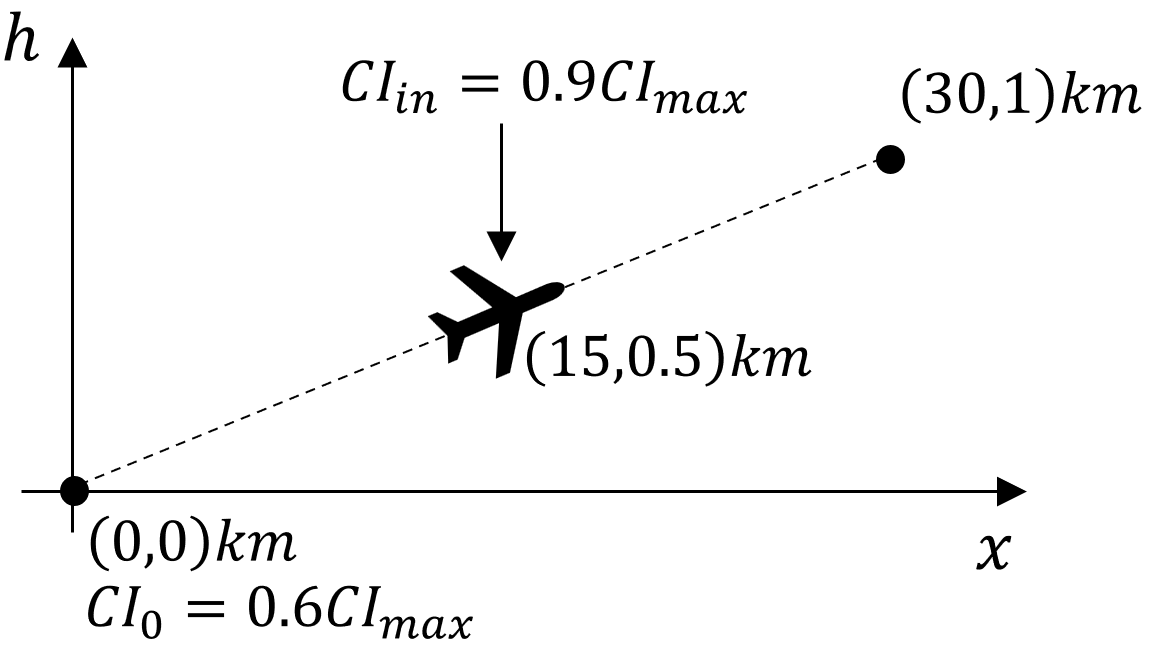}}
\caption{Flight scenario}
\label{fig_sce}
\end{figure}

This section presents a simulated scenario involving an all-electric aircraft during climb, where the optimal airspeed and climbing time are determined. The optimal aircraft{'}s airspeed values were computed in MATLAB by the \textit{fzero} function using (\ref{eq_dJ}) and (\ref{eq_v0}) and the climbing time was found by solving (\ref{eq_tc}). In this scenario, the aircraft departs from the origin waypoint $(x_{0},h_{0}) = (0,0) km$ and reaches the waypoint that indicates the beginning of cruise, noted as $(x_{c},h_{c}) = (30,1) km$. The initial cost index $CI_{0} = 0.6CI_{max}$ was defined by the airline, assuming $0 \leq CI \leq CI_{max}$, where $CI_{max}$ corresponds to the maximum value of $CI$, with the aircraft operating within its envelope, as per assumption 1). The optimal climb airspeed $v_{0}^{*}$ was computed using (\ref{eq_v0}) with $CI_{0}$. The climbing time, which in this case is the scheduled climbing time $t_{c_{0}}^{*}$ is found using (\ref{eq_tc}). This scenario is inspired by operational procedures that require constant airspeed in climb, such as the NADPs, adapted for an all-electric aircraft. In this sense, ATC imposes a change in the aircraft{'}s airspeed while it is climbing at the intermediate position $(x_{int},h_{int}) = (15,0.5) km$ after time $t_{1}$ has passed, to comply with the local noise abatement mandates by providing the cost index $CI_{in} = 0.9 CI_{max}$. The adjusted optimal airspeed in the second climb segment $v^{*}_{1}$ and the climbing time $t_{c_{1}}^{*}$ will be then computed using (\ref{eq_dJ}) and (\ref{eq_tc}), respectively, with $(x_{0},h_{0}) = (x_{int},h_{int})$.
Figure \ref{fig_sce} summarizes the described flight scenario.

The average value of the climb rate $\bar{\dot{h}}$ is determined based on operational procedures or past data. If we consider the example of a departure from the Pierre Elliot Trudeau International
Airport runway 06L for a reference ground speed of $75$ $knots$, the climb rate that complies with the minimum climb gradient is $325 ft/NM$, which is equivalent to $1.65 m/s$ \cite{Jeppesen_YUL2013}. Therefore, for the simulated scenario, $\bar{\dot{h}} = 1.65 m/s$.

In the numerical simulation, the values of $\bar{\rho}$ and $\bar{\delta_{\rho}}$ were computed using (\ref{eq_rho_bar}) and (\ref{eq_delta_rho_bar}), respectively, considering that the altitude of the aircraft is a discrete-time function $H$ bounded by the initial altitude $h_{0}$ and final altitude $h_{c}$ and (\ref{rho_troposphere}) that correlates the air density with the altitude.

\begin{equation}
     \bar{\rho} = \frac{1}{h_{c}-h_{0}} \sum_{H=h_{0}}^{h_{c}} \resizebox{.55\hsize}{!}{$4.1748e^{-11}(288.14 - 0.00649H)^{4.256}$}\label{eq_rho_bar}
\end{equation}

\begin{equation}
     \bar{\delta_{\rho}} = \frac{1}{h_{c}-h_{0}} \sum_{H=h_{0}}^{h_{c}} \frac{1}{\resizebox{.55\hsize}{!}{$4.1748e^{-11}(288.14 - 0.00649H)^{4.256}$}}\label{eq_delta_rho_bar}
\end{equation}

\begin{figure}[ht!]
\centerline{\includegraphics[scale=0.41]{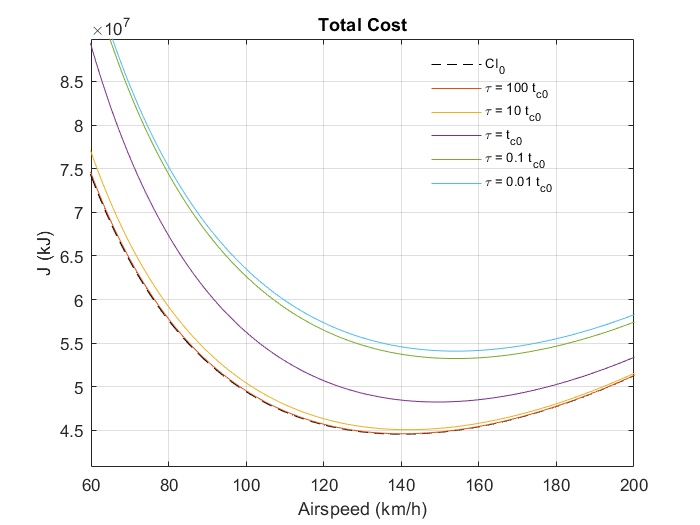}}
\caption{Total Cost as a function of the aircraft’s airspeed}
\label{fig_cost}
\end{figure}

As discussed in \cite{Silva_Akgunduz_Rodrigues2024}, the time constant $\tau$ of the first-order filter that models the time-varying cost index determines how fast $CI$ converges to the commanded value $CI_{in}$. Figure \ref{fig_cost} shows the total cost $J$ as a function of the aircraft airspeed for different values of $\tau$. The dashed line represents the total cost function for an aircraft operating with a constant $CI=CI_{0}$. 

Smaller values of $\tau$ enable $CI$ to reach the mandated value set by ATC faster than higher values of $\tau$, thereby mitigating the risk of non-compliance with operational regulations. In cases where $CI_{in}>CI_{0}$, the optimal climb airspeed for smaller values of $\tau$ is higher than the optimal climb airspeed computed for larger values of $\tau$. As a consequence, the total energy consumption is also higher for aircraft operating with smaller values of $\tau$. Based on the observed behavior of $CI$ for different values of the time constant, a value of $\tau = 0.01t_{c_{0}}^{*}$, was chosen as the first-order filter parameter considered herein.

\begin{figure}[hbt!]
\centerline{\includegraphics[scale=0.60]{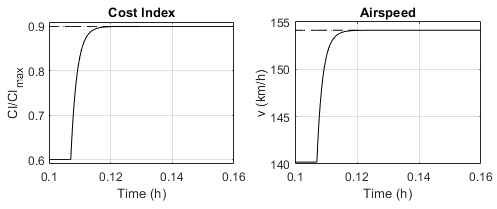}}
\caption{Cost index (left) and airspeed (right) as a function of climbing time}
\label{fig_CI_v}
\end{figure}

Figure \ref{fig_CI_v} depicts the time-varying cost index (left) and the aircraft airspeed (right) as a function of time. The increase in $CI$ also increases the aircraft airspeed. The parameter $\tau$ is chosen in such a way that $CI$ converges fast to the commanded value $CI_{in}$ and the aircraft{'}s airspeed also rapidly transitions to the optimal solution that accommodates the ATC input.

\begin{figure}[hbt!]
\centerline{\includegraphics[scale=0.34]{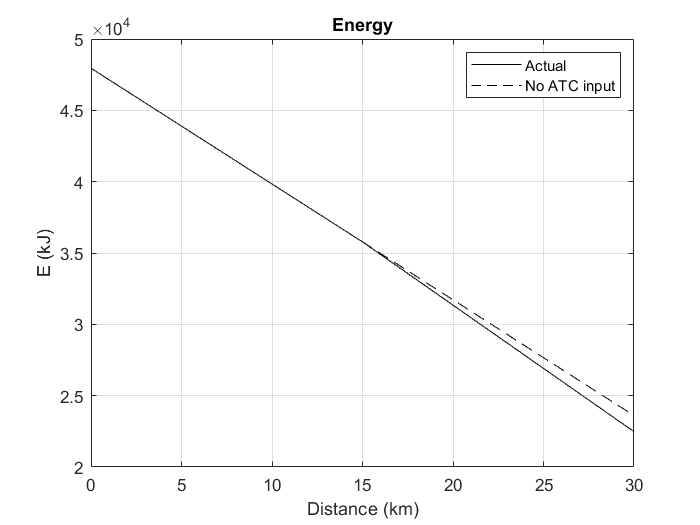}}
\caption{Available energy as a function of distance travelled}
\label{fig_energy}
\end{figure}

The energy consumption also increases during climb for a higher airspeed. In Figure \ref{fig_energy}, the available energy is depicted as a solid line as a function of the distance travelled, whereas the dashed line represents the available energy if the aircraft operated as per its original schedule, with no ATC input. However, the increase in airspeed to comply with the mandated $CI_{in}$ caused an increase in the energy consumption, represented by a smaller value in the final available energy, compared to the dashed line. 

In summary, for the simulated scenario,  $v_{0}^{*} = 140.19 km/h$ and $t_{c_{0}}^{*} = 12min51s$. At $t_{1} = 6min26s$, the ATC input was received. Then, the optimal airspeed was adjusted to $v^{*}_{1} = 154.13 km/h$, leading to $t_{c_{1}}^{*} = 12min16s$, which results in a shortening of $35s$ in the climb phase duration.

\section{Conclusions}\label{sec5}
This paper introduces a novel methodology for calculating constant airspeed and flight time for all-electric aircraft during climb, incorporating a time-varying cost index. The approach can be utilized across various operational procedures that necessitate climbing at a constant airspeed to meet ATC regulations. Validation through a simulated scenario showed that the optimal climb airspeed and climbing time for an all-electric aircraft are influenced by the time constant of the first-order filter used to model the variable CI, as well as the cost index set by ATC. The optimal values for airspeed, climbing time, and energy consumption were determined for an all-electric aircraft, paving the way for applying this methodology to future advanced air mobility all-electric vehicles.

\printbibliography
\end {document}